\documentclass[12pt,preprint]{aastex}
 
\begin{document}

\title{Ultraviolet spectroscopy of narrow coronal mass ejections}
 
\author{D. Dobrzycka\altaffilmark{1}, J. C. Raymond\altaffilmark{1},
D. A. Biesecker\altaffilmark{2}, J. Li\altaffilmark{3} and A. Ciaravella\altaffilmark{4}}

\altaffiltext{1}{Harvard-Smithsonian Center for Astrophysics,
60 Garden Street, Cambridge MA 02138, USA,
[ddobrzycka,jraymond]@cfa.harvard.edu}

\altaffiltext{2}{NOAA/Space Environment Center,
325 Broadway, Boulder CO 80305, USA, Doug.Biesecker@noaa.gov}

\altaffiltext{3}{Institute for Astronomy,
2680 Woodlawn Drive, Honolulu HI 96822, USA, jing@ifa.hawaii.edu}

\altaffiltext{4}{Osservatorio Astronomico di Palermo ``G. S. Vaiana'',
Piazza Parlamento 1, 90134 Palermo, Italy, ciarave@astropa.unipa.it}
 
\shorttitle{UV spectroscopy of narrow CMEs}
\shortauthors{Dobrzycka et al.}

\slugcomment{Accepted for publication in Ap.J.}


\begin{abstract}

We present Ultraviolet Coronagraph Spectrometer (UVCS) observations of
5 narrow coronal mass ejections (CMEs) that were among 15 narrow CMEs
originally selected by Gilbert et al. (2001). Two events (1999 March
27, April 15) were ``structured'', i.e. in white light data they
exhibited well defined interior features, and three (1999 May 9, May
21, June 3) were ``unstructured'', i.e. appeared featureless. In UVCS
data the events were seen as $4-13^\circ$ wide enhancements of the
strongest coronal lines \ion{H}{1} Ly$\alpha$ and \ion{O}{6}
$\lambda\lambda1032,1037$. We derived electron densities for several
of the events from the Large Angle Spectrometric Coronagraph (LASCO)
C2 white light observations. They are comparable to or smaller than
densities inferred for other CMEs. We modeled the observable
properties of examples of the structured (1999 April 15) and
unstructured (1999 May 9) narrow CMEs at different heights in the
corona between $1.5~R_\odot$ and $2~R_\odot$. The derived electron
temperatures, densities and outflow speeds are similar for those two
types of ejections. They were compared with properties of polar
coronal jets and other CMEs. We discuss different scenarios of narrow
CME formation either as a jet formed by reconnection onto open field
lines or CME ejected by expansion of closed field structures. Overall,
we conclude that the existing observations do not definitively place
the narrow CMEs into the jet or the CME picture, but the acceleration
of the 1999 April 15 event resembles acceleration seen in many CMEs,
rather than constant speeds or deceleration observed in jets.

\end{abstract}

\section{Introduction}

Coronal mass ejections (CMEs) are dynamic solar phenomena in which
mass and magnetic field are ejected from the lower corona into the
higher solar atmosphere and interplanetary space. They involve large
scale changes to the coronal structure and reconfiguration of the
coronal magnetic field. A commonly used definition of CMEs describes
them as new, discrete, bright features appearing in the field of view
of a white light coronagraph and moving outward over a period of
minutes to hours (e.g. Munro et al., 1979). On the basis of this broad
definition several classifications were introduced.  The morphology
and spectral characteristics of CMEs were extensively studied over the
last decade (e.g. Hundhausen 1997; St. Cyr et al. 2000; Antonucci et
al. 1997; Ciaravella et al. 1997, 2000, 2001).  Apparent angular
widths of the CMEs cover a wide range from few to 360 degrees (Howard
et al. 1985; St. Cyr et al. 1999, 2000).  The very narrow structures
(narrower than about $15^\circ - 20^\circ$) form only a small subset
of all the observed CMEs and are usually referred to as rays, spikes,
fans, etc. (Munro \& Sime 1985; Howard et al. 1985).

Recently, Gilbert et al. (2001) conducted a study of 15 selected
narrow CMEs observed in white light by the Large Angle Spectrometric
Coronagraph (LASCO) aboard {\it Solar and Heliospheric Observatory}
({\it SOHO}) in the period from 1999 March through December. The
selection criteria included the condition that the measurable angular
width of the event is no more than $15^\circ$ and the CME is not part
of a larger event. Moreover, the CMEs had to have a clear surface
association in H$\alpha$ and/or \ion{He}{1} data from the Mauna Loa
Solar Observatory (MLSO), or in data from the Extreme Ultraviolet
Imaging Telescope (EIT) on {\it SOHO}. Gilbert et al. (2001) examined
the events' structures, angular sizes and projected radial velocities
within the LASCO C2 and C3 coronagraphs field of view
($2-32~R_\odot$), and likely surface associations. The narrow CMEs
appeared to fall into two categories: ``structured'', that exhibit
well defined interior features in the LASCO images (7 events) and
``unstructured'', that were featureless (8 events). In their
exploratory study they did not find any obvious difference between
narrow and large CMEs other than their appearance. The projected
average radial velocities of the narrow CMEs ($\sim 400$~km~s$^{-1}$)
are similar to the average speed of the regular CMEs.

The solar corona manifests its activity in various ways. Besides CMEs,
another type of eruptive event observed throughout the whole solar
cycle is coronal jets. They are presumably triggered by field
reconnection between a magnetic dipole and neighboring unipolar region
(Wang et al. 1998). Different types of coronal jets were observed by
various instruments: the {\it Yohkoh} Soft X--Ray Telescope (SXT)
(Shibata et al. 1992) and {\it SOHO} Ultraviolet Coronagraph
Spectrometer (UVCS), EIT, and LASCO (Moses et al. 1997, Gurman et
al. 1998, Wang et al. 1998, Wood et al. 1999, Dobrzycka et al. 2000,
2002). The jets originate in a variety of settings; in UV and X-ray
bright points within coronal holes, in active regions, flares, etc. In
the white light coronagraphs they appear as narrow, fast and
featureless, bright structures (Howard et al. 1985). They are often
listed among CMEs as they satisfy the general CME definition.

According to the most popular models, coronal jets and CMEs are
entirely different phenomena.  A jet is taken to be the result of
reconnection between open and closed field lines in which the gas may
be strongly heated (X--ray jet), and it is accelerated by the magnetic
tension force of the newly reconnected field lines (Shibata et
al. 1992; Shimojo et al. 1996). CMEs, on the other hand, are
attributed to the release of magnetic stress stored in a twisted flux
rope (e.g. Amari et al. 2000; Gibson \& Low 1998; Lin \& Forbes 2000)
or in a sheared arcade held down by overlying unsheared field
(Antiochos, DeVore, \& Klimchuk 1999).

In this project we study spectra of the narrow CMEs as they may shed
some light on differences between narrow and large CMEs and, in
particular, the relation of the narrow CMEs to coronal jets. We
present ultraviolet spectroscopy of the several narrow CMEs that were
originally selected by Gilbert et al. (2001). The data were obtained
with the UVCS instrument on board {\it SOHO}. We
describe the observations in {\S} 2. In {\S} 3 we discuss our results,
model the observations, and compare narrow CMEs with coronal jets and
other CMEs. Finally, we summarize the results in {\S} 4.

\section{Observations}

The UVCS/{\it SOHO} instrument is designed to observe the extended
solar corona from $1.4~R_\odot$ to about $10~R_\odot$ (Kohl et
al. 1995). In 1999 the coronal spectra were acquired in the \ion{O}{6}
spectrometer channel, which covers the wavelength ranges
940--1123~\AA\ in first and 473--561~\AA\ in second order. A mirror
between the grating and the detector images spectral lines in the
range 1160--1270~\AA\ (580--635~\AA\ in second order).  The primary
lines observed were \ion{H}{1} Ly$\alpha$ $\lambda$1216 and \ion{O}{6}
$\lambda\lambda$ 1032,1037.

To obtain spectra of the narrow CMEs we searched the UVCS archival
data following the list of the LASCO events compiled by Gilbert et
al. (2001). In 1999, each UVCS daily observing plan consisted of a 14
hour standard synoptic sequence and special observations designed by
the lead observer. During the synoptic observations the slits were
scanned radially at eight position angles (P.A.; measured
counterclockwise from the north heliographic pole) $45^\circ$ apart at
different heights ranging from $1.4~R_\odot$ up to $3.6~R_\odot$. UVCS
had a chance to record the CME spectrum only when its pointing
coincided with the event. This happened on five occasions out of
fifteen listed by Gilbert et al. (2001). Most of the observations
appeared as a part of the synoptic sequence.  During the synoptic
scans the instrument configuration included an entrance slit of 100
\micron\ for the \ion{O}{6} channel corresponding to 28'' or
0.4~\AA. The spatial binning was 3 pixels (21''), and the grating
position covered three spectral ranges: 1023.20--1043.75,
1208.72--1220.62 (with \ion{H}{1} Ly$\alpha$ from the redundant
mirror), 975.78--979.16 {\AA}, with spectral binnings of 3, 2, and 2
pixels (0.298, 0.183, and 0.183 {\AA}) respectively. The exposure time
was $180-200$~s. Unfortunately, the synoptic program provides fairly
short dwell times at each height, making it difficult to characterize
time dependence of the events.

We obtained ultraviolet spectra of five narrow CMEs reported by
Gilbert et al. (2001). The events on 1999 March 27 and April 15 were
described as structured and events on 1999 May 9, 21, and 1999 June 3
as unstructured. Table~1 contains a summary of observations.

In the data reduction we followed the standard techniques described in
Kohl et al. (1997, 1999). We used the latest version of the UVCS Data
Analysis Software (DAS) (released in June 2001) for wavelength,
intensity calibration, and removal of image distortion. The
uncertainties in the line intensities are estimated to be $\pm20\%$
and are due to photon--counting statistics, background subtraction,
and radiometric calibration (Gardner et al. 2002). Table~2 lists
measured profiles of the brightest emission lines, \ion{H}{1}
Ly$\alpha$, \ion{O}{6} $\lambda\lambda1032,1037$ after the background
corona (pre-- or post--CME spectra, see below) was subtracted. In most
cases (except very weak lines) to obtain the total intensities and
$1/e$ widths we fitted multiple Gaussian functions to the observed
profiles. A detailed description of this technique including the
removal of the instrumental profile, background, stray light and
interplanetary hydrogen contributions can be found in Kohl et
al. (1997).

Following Gilbert et al. (2001) we attempted to identify the solar
disk and/or lower corona associations of these events. We used data
from EIT and SXT instruments. Also, most of the events were described
in the daily reports by EIT and LASCO teams\footnote{Reports available
from http://sohowww.nascom.nasa.gov/}.

\subsection{1999 March 27 (structured)}

Gilbert et al. (2001) reported that this CME appeared in the LASCO C2
field of view at P.A$=310^\circ$, 16:54 UT and was $12^\circ$ wide.
UVCS began a special CME watch campaign at 16:02 UT and continued it
until 00:12 UT of the next day. The observations were centered at
P.A.$=310^\circ$ and consisted of repeating sequences of eight
exposures at $1.49~R_\odot$ and two at $1.85~R_\odot$. Each exposure
was 75~s. The instrument configuration included an entrance slit of
$50\micron$ and spatial binning of 3 pixels (21''). The grating and
detector mask covered four spectral ranges: $1030.45-1039.16$,
$997.53-1000.51$, $1212.14-1216.71$, $976.13-978.91$~\AA, which
allowed us to monitor the emission lines \ion{H}{1} Ly$\alpha$,
\ion{O}{6} $\lambda\lambda1032, 1037$, \ion{C}{3} $\lambda977$,
\ion{Si}{12} $\lambda499$ in second order, and others. The binning in
the spectral direction was 2, 3, 1, and 2 pixels (0.199, 0.298, 0.092,
0.199~\AA) in the four ranges, respectively.

The evidence for the CME passing the UVCS slit was seen from the
beginning of observations at $1.49~R_\odot$. It appeared as an
enhancement of the Ly$\alpha$ and \ion{O}{6} $\lambda\lambda1032,1037$
line intensities in the spatial range between -100 and 250 arcsec,
which corresponds to about $13^\circ$ of angular width. At 16:29 UT
the slit was moved to $1.85~R_\odot$ for 3 minutes and the line
intensity enhancement there was visible between -100 and 215 arcsec,
which is consistent with $\sim 10^\circ$ of angular width. During the
long coverage at $1.49~R_\odot$ both \ion{H}{1} Ly$\alpha$ and
\ion{O}{6} lines showed similar behavior. They reached maximum at
16:05--16:12 UT, then the intensity slightly decreased for about 10
minutes and increased again but did not reach the initial value. From
about 17:25 UT the line intensities became consistent with the
background.

We compared the CME observations with corresponding data obtained
within the synoptic program on the same day but before the event. The
radial scan centered at P.A.=315$^\circ$ included observations at
$1.49~R_\odot$ taken between 03:36 -- 03:43 UT and observations at
$1.85~R_\odot$ between 03:54 -- 04:04 UT. We considered the correction
for different pointing angle and verified that the same region in the
pre--CME and CME data that was away from the CME remained
unchanged. The emission line intensity variations away from the CME
were within $10-15$\% and $1/e$ widths were comparable at both
heights. With respect to the pre--CME data the maximum \ion{H}{1}
Ly$\alpha$ intensity during the event was enhanced by $30-40$\% at
both heights, while the \ion{O}{6} lines showed increase by about 20\%
and close to 100\% at $1.49~R_\odot$ and $1.85~R_\odot$
respectively. The measured line intensities with subtracted pre--CME
profiles are summarized in Table~2. The \ion{O}{6} $\lambda1032$ and
\ion{O}{6} $\lambda1037$ intensity ratio was 3 before the CME. During
the ejection it dropped to about 2 at both heights due to the Doppler
dimming effect (see {\S} 3). The $1/e$ widths of the emission line
profiles appeared comparable to the pre--CME values. In the post--CME
spectra the lines were weaker than during the CME passage but still
brighter than before the event. We did not detect any obvious Doppler
shifts. Both \ion{C}{3} and \ion{Si}{12} line intensities varied
throughout the observations but the profiles were weak and could not
be fitted. Also it was not clear if the variation was due to the CME.

On the disk, EIT data appeared to show the CME associated with a
dark filament. It began to bubble between 12:00 and 15:36 UT and at
15:36 UT a dimming typical for CMEs was visible behind the limb.
There were no SXT images available between 12:25 and 17:13 UT. After
that the SXT data did not show any obvious association up to 18:43
UT. However, the next available image at 19:57 UT clearly displayed
loops formed in that area.

\subsection{1999 Apr 15 (structured)}

This event was recorded by LASCO C2 at 07:54 UT, P.A.$=102^\circ$ and
appeared to be $10^\circ$ wide. UVCS was taking a synoptic radial scan
at P.A.$=90^\circ$ between 07:33 and 09:48 UT. The CME was detected at
all heights: $1.38, 1.50, 1.62, 1.74, 1.98, 2.33, 2.88,$ and
$3.62~R_\odot$. Brightening of the \ion{H}{1} Ly$\alpha$ and both
\ion{O}{6} $\lambda\lambda1032,1037$ lines due to the CME was present
at all exposures. However, during observations at $3.62~R_\odot$ the
brightening was initially weak, becoming obvious only after several
exposures, at about 09:31 UT. Then, the \ion{H}{1} Ly$\alpha$
intensity increased by 35\%. The \ion{O}{6} lines were too weak at
this height and we were unable to fit the profiles. The bright feature
in the CME spectra was centered at P.A.$=107-108^\circ$ and was
$5-7^\circ$ wide. Figure~1 shows an example of spectral lines
obtained at three different heights. The line profiles are blue
shifted by about 50~km~s$^{-1}$.

\begin{figure}
\epsscale{0.91}
\plotone{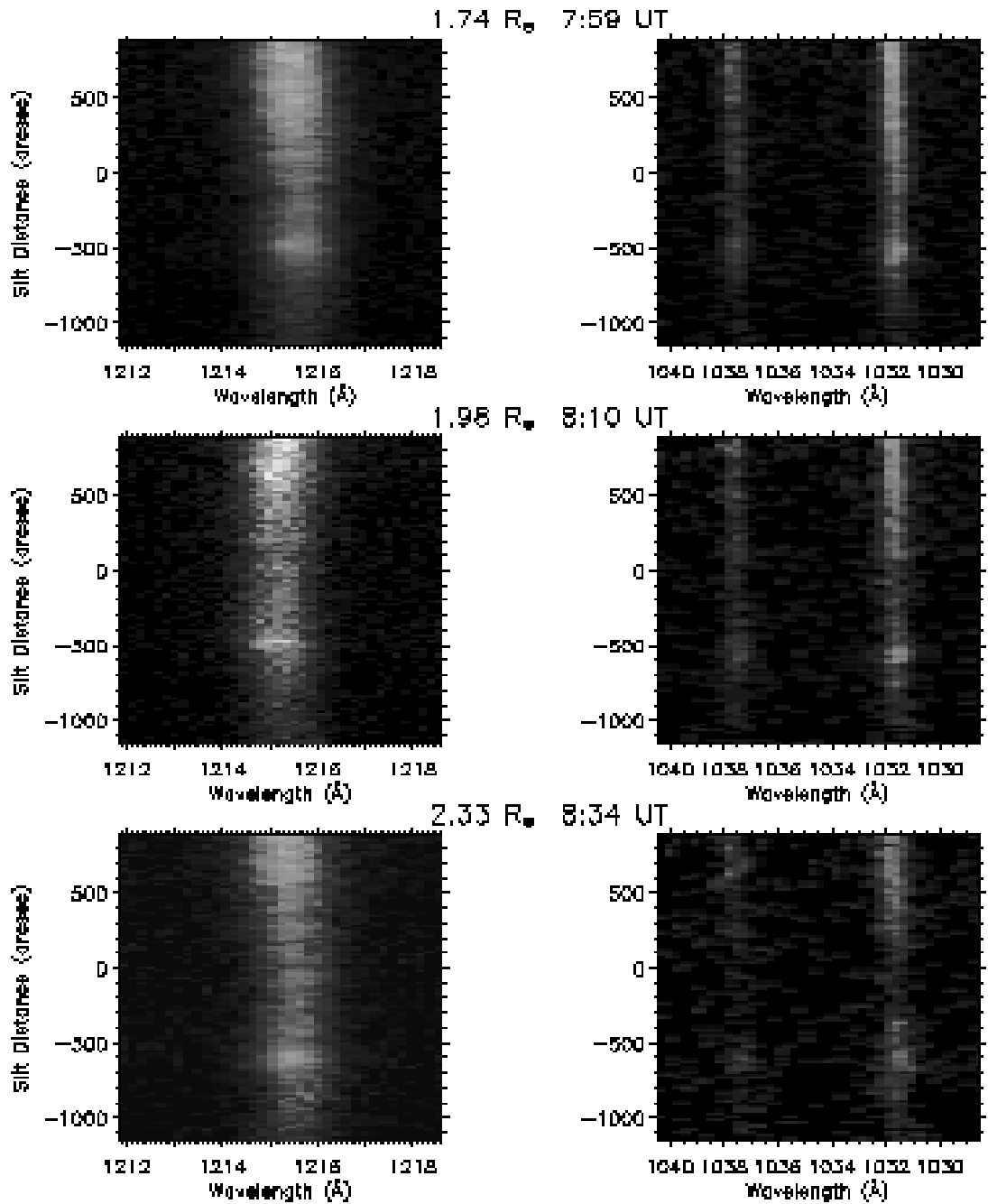}
\caption{Examples of the UVCS observations of the 1999 April 15
narrow CME. The projected height and initial UT time of each exposure
are shown. The vertical axis is the spatial coordinate along the UVCS
entrance slit (in arcsec), centered at P.A.$=90^\circ$ at the
indicated heliocentric distance. The CME appeared as a narrow
brightening in the \ion{H}{1} Ly$\alpha$ $\lambda1216$ ({\it left})
and \ion{O}{6} $\lambda\lambda1032,1037$ ({\it right}) emission lines
at about $-600$ arcsec.}
\end{figure}

To compare the CME line profiles with the background corona we looked
at the corresponding synoptic scan obtained a day earlier. However, it
contained another transient structure visible at most heights and the
line intensities were enhanced with respect to the background
level. We found that the synoptic observations obtained 24 hours after
the CME, on 1999 April 16, are more suitable for comparison. The line
intensities, in the region of the corona away from the CME, were
consistent with each other within $10-15$\% and the $1/e$ widths were
comparable at all heights. With respect to those data, \ion{H}{1}
Ly$\alpha$ line brightening due to CME ranged from 25\% at
1.38~$R_\odot$ to about 50\% at $1.98~R_\odot$, while the \ion{O}{6}
lines were enhanced by $100-350$\% at different heights. The
\ion{O}{6} $\lambda\lambda1032, 1037$ line ratio was reduced from 2.8
to about $2-2.5$. The measured line intensities, after the post--CME
profiles were subtracted, are summarized in Table~2.

Gilbert et al. (2001) noted that the EIT data point to association of
this CME with a flare in the active region close to the East limb. In
running difference EIT $\lambda195$ images the activity appears to
start between 05:36 and 06:00 UT with brightening and rising of
several arcs inside the active region. By 08:12 UT the loop legs
extend up reaching the edge of EIT field of view.
There are no SXT images available for that period of time. In
the LASCO daily report the event was described as a jet--like
ejection.

\subsection{1999 May 9 (unstructured)}

This narrow CME appeared in LASCO C2 at 19:30 UT, P.A.=$100^\circ$
(Gilbert et al. 2001). It was recorded by UVCS in the extended
synoptic observations at P.A.$=90^\circ$ between 19:17 and 20:53
UT. Brightening of the \ion{H}{1} Ly$\alpha$ and \ion{O}{6}
$\lambda1032,1037$ lines centered at about P.A.$=100^\circ$ was
visible at 1.38, 1.58, and $1.86~R_\odot$. The \ion{O}{6} lines showed
enhancement also at $2.15~R_\odot$. We did not measure any significant
Doppler shifts of the line profiles. The angular width of the CME was
$4^\circ$.

To compare the CME line intensities with pre--CME profiles we used the
corresponding synoptic data from the previous day, 1999, May 8. They
were centered at the same position angle -- P.A.$=90^{\circ}$, and
covered heights from 1.38 to $3.62~R_\odot$ between 09:01 UT and 11:16
UT. We compared the emission line intensities in the region of corona
away from the CME. The difference between line intensities on 1999,
May 8 and May 9 was no more than $10-15$\% and thus we accepted it as
satisfactory agreement. The CME \ion{H}{1} Ly$\alpha$ was $15-30$\%
and \ion{O}{6} $\lambda1032,1037$ $20-50$\% brighter with respect to
the 1999, May 8 observations. The $1/e$ widths did not change
significantly. The \ion{O}{6} line ratio decreased to 2.0 compared
with the pre--CME value of about 3. Table~2 includes a summary of the
measured line intensities after the pre--CME spectrum was subtracted.

Gilbert et al. (2001) noted that the CME was associated with an active
prominence. It most likely originated from the back of the disk. The
SXT image from 22:15 UT does not show any significant features that
could be associated with the CME.

\subsection{1999 May 21 (unstructured)}

At the time of the CME passage UVCS was executing the synoptic
program; the radial scan centered at P.A.$=270^\circ$ in
particular. Gilbert et al. (2001) reported 02:06 UT as the time of
appearance of the CME in the C2 field of view at
P.A.$=290^{\circ}$. UVCS was observing at $3.62~R_\odot$ from 01:58 UT
to 02:36 UT and recorded brightening of the \ion{O}{6}
$\lambda1032,1037$ lines beginning at 02:08 UT. The line brightening
occurred close to the edge of the UVCS slit in the region
corresponding to P.A.$\approx 289^\circ$. We did not see any
significant intensity variations in the \ion{H}{1} Ly$\alpha$ line.
Poor spectral resolution and the relative weakness of the \ion{O}{6}
lines so high in the corona prevented us from fitting the
profiles. However, we estimated that the integrated intensities
increased by more than $\sim 50$\% due to CME passage.

Gilbert et al. (2001) concluded that the CME was associated with a
flare in an active region located at the West limb.  The SXT image
from 02:44 UT shows a bright loop formed at this position.

\subsection{1999 June 3  (unstructured)}

This CME was first observed by LASCO C2 at 07:26 UT. UVCS began the
synoptic radial scan centered at P.A.$=90^\circ$ at 09:06 UT when the
CME seemed to be already fading away.  We detected narrow brightening
of \ion{H}{1} Ly$\alpha$ and \ion{O}{6} $\lambda 1032,1037$ lines in
exposures between 10:15 and 10:42 UT at $2.88~R_\odot$ (see Table~2
for measured line intensities). The structure appeared at
P.A.=$86^\circ$ and its angular width was $4^\circ$. There is a
discrepancy with Gilbert et al. (2001). They noted that the CME was
centered at P.A.=$90^\circ$ rather than $86^\circ$, however we
carefully analyzed the LASCO C2 and C3 movies from that period of time
and concluded that the CME was really seen a few degrees north of the
P.A.=$90^\circ$ mark. Also the angular width measured with UVCS agrees
with LASCO estimates, which suggests that it is the same structure.
There were no other ejections in the close vicinity.

We compared the CME spectra with corresponding synoptic observations
obtained a day later an 1999 June 04 between 09:03 UT and 11:18
UT. Synoptic data acquired a day earlier contained some transient
features so we could not use them. At $2.88~R_\odot$, away from the
CME position, the integrated intensities of the brighter emission
lines agreed with those in the CME spectra within $5-10$\%.  With
respect to the post--CME data the passage of the CME caused $75-100$\%
enhancement of the integrated intensities of \ion{O}{6}
$\lambda1032,1037$ and 12\% increase in the \ion{H}{1} Ly$\alpha$
intensities. The \ion{O}{6} line ratio decreased from 3.3 to 2.6. We
did not see any significant changes in the $1/e$ widths of the
profiles. However, the lines were relatively weak so high in the
corona and the \ion{O}{6} profiles were heavily binned in the spectral
direction ($\delta\lambda \approx 0.3$\AA), which prevented us from
measuring the $1/e$ widths accurately.

Gilbert et al. (2001) found that this CME was likely lifted up from a
surface in the area without a preexisting structure and they
classified this event as a ``surge''. At 05:53 UT SXT observed a faint
jet--like structure on the East limb. Later, at 07:31 UT the
brightness increased in this area that appeared to be connected to the
nearby active region by faint jet--like features.

\section{Discussion}

UVCS recorded five out of 15 narrow CMEs selected by Gilbert et
al. (2001). Two of them were structured, i.e. in white light data they
exhibited well defined interior features, and three unstructured,
i.e. appeared featureless. In our studies we relied on the archival
UVCS data obtained when the instrument pointing coincided with
occurrence of the CME. Whenever UVCS was pointing in the right
direction at the right time, the CME was recorded as a brightening of
the strongest emission lines.  In most cases the configuration of the
instrument and type of UVCS observations was not ideal for a
CME--watch campaign, e.g., heavy binning in the spectral direction for
the \ion{O}{6} lines, short time coverage at each height, etc.

The narrow CMEs appear to be associated with disk activity that lasted
up to several hours. In the LASCO C2 field of view they were visible
for an extended period of time of about $1-1.5$ hour.  From the time
of first appearance in C2, estimated speeds and recorded beginnings
for some events we were able to reconstruct the timeline of the
considered CMEs assuming constant velocity.  We found that in all
events the UVCS observations at each height began from 10 minutes to
about an hour after the passage of the CME leading edge. Thus, in most
cases UVCS recorded only the trailing parts of the CME. Also at
different heights UVCS most likely probed different parts of the CME.

We obtained LASCO C2 mass estimates for most of the events. They were
derived from the brightness enhancement in white light data due to the
CME after subtracting a pre--event image. The number of coronal
electrons needed to produce the excess brightness was calculated
assuming Thompson scattering of photospheric light. The mass was then
computed assuming a neutral atmosphere consisting of ionized hydrogen
and 10\% helium (for a more comprehensive discussion of mass
calculations with LASCO, see Vourlidas et al. 2000). We assumed that
all of the material contributing to the excess mass lies in the plane
of the sky, which is consistent with mostly negligible blue and
redshifts of their profiles. To derive the density we assumed that the
narrow CMEs can be approximated with a cylindrical geometry within the
LASCO C2 field of view. The length of the cylinder was determined from
the radial length of the CME in C2 images and a width consistent with
the observed angular size. This approach assumes that the density of
material in the structure is uniform, which is likely an
oversimplification. As each narrow CME was visible in LASCO C2 field
of view for about an hour, several images were taken at this period of
time. We derived mass and density estimates for all the frames
containing the CME structure and adopted an average value. Table~3
lists the derived densities, the average height the structure appeared
at and the observed width.

Wang \& Sheeley (2002) examined a number of jetlike events observed
with LASCO during the current sunspot maximum. They appear to be
similar to the narrow CMEs -- their angular widths are $\sim 3^\circ -
7^\circ$ and have clear surface association visible in the EIT
$\lambda195$ data. Many of the ejections originated from active
regions located inside or near the boundaries of nonpolar coronal
holes. Based on that Wang \& Sheeley concluded that the jet--producing
regions consist of systems of closed magnetic loops and adjacent open
flux. Perturbation in the closed field would cause the lines to
reconnect with the overlying open flux, releasing the trapped energy
in the form of a jetlike ejection. We examined the \ion{He}{1}
$\lambda10830$ spectroheliograms and \ion{Fe}{1} $\lambda8688$
magnetograms obtained by the National Solar Observatory Vacum
Telescope at Kitt Peak as well as magnetograms taken with the {\it
SOHO} Michelson Doppler Imager (MDI) corresponding to ejections in our
sample. In the events on May 9 and May 21 there was a nonpolar coronal
hole located close to the active region that was presumably associated
with the narrow CME. On March 27 and June 6 there were small coronal
holes within $10-20^\circ$ of latitude from what was interpreted as
footpoints of the CME. There was no obvious coronal hole present close
to the source of the April 15 ejection.

To model narrow CME plasma conditions we used a time--independent
spectral line synthesis model based on the code developed by Cranmer
et al. (1999). The model describes the narrow CME as a flux tube
located in the plain of the sky, which is consistent with the lack of
significant Doppler shifts of the emission lines (see also Dobrzycka
et al. 2000, 2002). There is no line--of--sight contribution from the
background and foreground corona taken into account. The goal of the
model is to reproduce the CME emission from Table~2 where the ambient
corona is already subtracted. The CME flux tube is defined at each
height to have the observed angular width, and the electron density
($n_e$), outflow speed, electron temperature ($T_e$), and ion
temperature (parallel, $T_\parallel$, and perpendicular, $T_\perp$, to
the radial directed field) are modeled .

We chose to model the best observed examples of structured (April 15)
and unstructured (May 09) narrow CMEs.  There were several
observational parameters to fit at several projected heights; the
\ion{H}{1} Ly$\alpha$ and \ion{O}{6} intensities, the $1/e$ widths of
the \ion{H}{1} Ly$\alpha$ and \ion {O}{6} profiles as well as the
\ion{O}{6} $\lambda\lambda1032,1037$ line ratio.

\subsection{1999 April 15 -- structured CME}

In our analysis of this event we concentrated on two heights:
$1.50~R_\odot$ and $1.98~R_\odot$.  Gilbert et al. (2001) estimated
that the leading edge or points close to the front of this CME moved
with constant velocity of 523~km~s$^{-1}$. They were unable to
identify acceleration or deceleration in this event. However,
independent analysis of the trajectory in the field of view of LASCO's
C2 and C3, available in the CSPSW/NRL {\it SOHO}/LASCO CME catalog
clearly revealed acceleration.  The \ion{O}{6}
$\lambda1032/\lambda1037$ line ratio is a good indicator of the plasma
outflow velocity (see e.g. Withbroe et al. 1982; Noci et
al. 1987). Figure~2 shows the model \ion{O}{6}
$\lambda1032/\lambda1037$ line ratios computed for the 1999 April 15
event at $1.50$ and $1.98~R_\odot$ for different values of $n_e$ and
$T_e$. Three sets of $n_e$ values were chosen to correspond to: (1)
the background coronal hole (Guhathakurta \& Holzer 1994), (2) the
background streamer (Guhathakurta \& Fisher 1995), and (3) a density
several times of that of the streamer. The model predictions are
overplotted with the observed ratios (1.9 at $1.50~R_\odot$ and 2.2 at
$1.98~R_\odot$) that have relatively large uncertainties due to
uncertain post--CME background subtraction and the \ion{O}{6} line
intensity variations from exposure to exposure. Figure~2 demonstrates
that if the electron density is comparable to or higher than the
streamer background corona we cannot identify a unique value for the
CME outflow velocity from the \ion{O}{6} ratio alone.

\begin{figure}
\epsscale{0.89}
\plotone{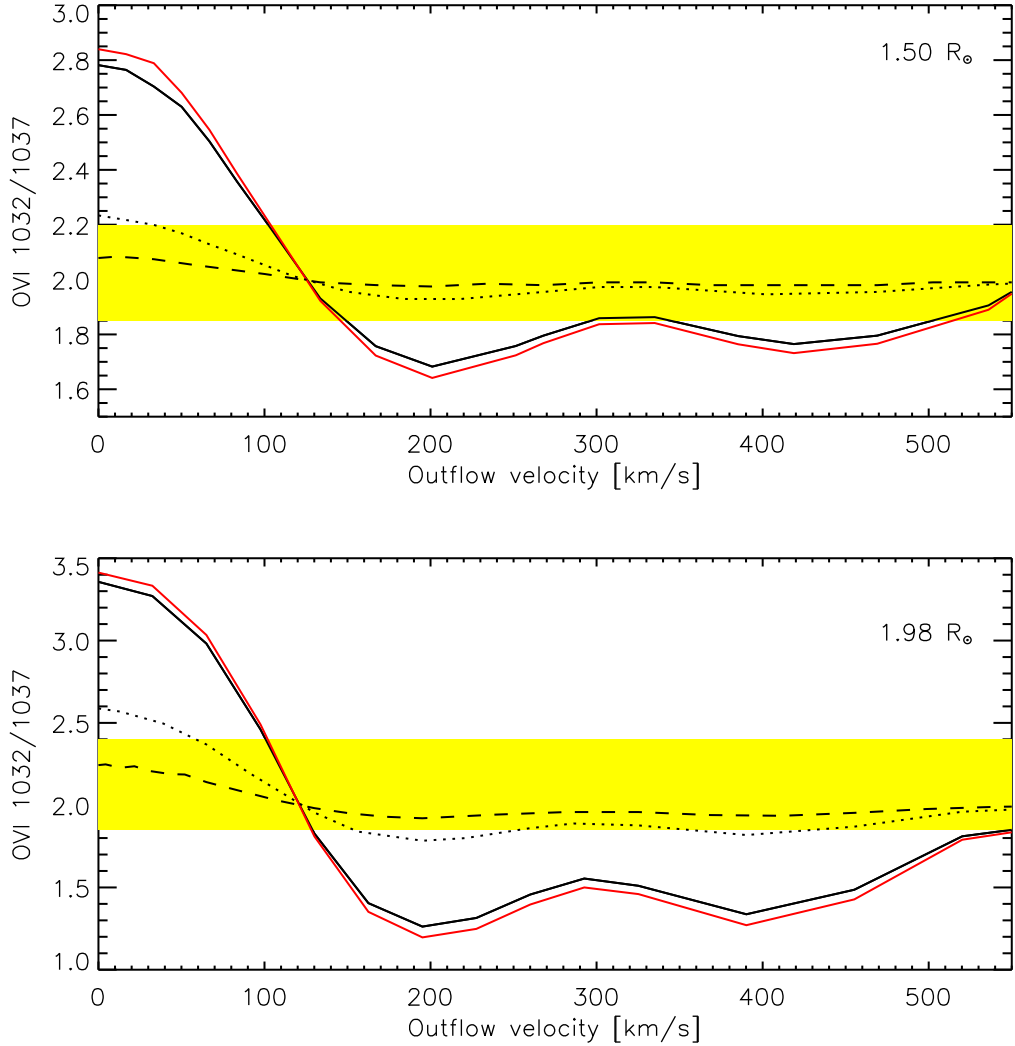}
\caption{Model \ion{O}{6} (1032,1037\AA) line ratio as a function of
outflow velocity (in km~s$^{-1}$) for the 1999 April 15 event. The
dependence is shown at two heights: $1.50~R_\odot ~(top~panel)$ and
$1.98~R_\odot ~(bottom~panel)$. The different black lines correspond
to different electron densities ($n_e$): solid -- $3.1\times10^6$ and
$4.6\times10^5$~cm$^{-3}$, dotted -- $1.6\times10^7$ and
$2.3\times10^6$~cm$^{-3}$, dashed -- $5.1\times10^7$ and
$7.0\times10^6$~cm$^{-3}$, for two heights respectively. We chose the
$n_e$ values so they correspond to the background coronal hole
(Guhathakurta \& Holzer 1994), background streamer (Guhathakurta \&
Fisher 1995) and several times that of the background streamer. The
\ion{O}{6} line ratios were computed using an electron temperature,
$T_e=5\times10^5$~K. The red lines correspond to the same $n_e$ as the
black lines but with $T_e=1.2\times10^6$~K. It shows that the ratios
do not depend significantly on the choice of $T_e$. The yellow block
corresponds to values allowed by observations.}
\end{figure}

At $1.50~R_\odot$ UVCS recorded strong relative brightening of the
\ion{O}{6} emission lines and only weak brightening of the \ion{H}{1}
Ly$\alpha$ (Table~2). It suggests that the \ion{H}{1} Ly$\alpha$
enhancement was most likely reduced due to Doppler dimming. Doppler
dimming arises from a Doppler shift between the incident chromospheric
radiation and the moving coronal gas. The faster the gas moves, the
larger the Doppler shift of its absorption profile, and thus fewer
photons are scattered to form the emission line. Doppler dimming
affects only the resonantly scattered part of the line. The coronal
\ion{H}{1} Ly$\alpha$ emission is strongly affected because it is
mostly resonantly scattered, while the \ion{O}{6} lines have
substantial collisional components. The model was able to reproduce
all the observational parameters: the \ion{H}{1} Ly$\alpha$ and \ion
{O}{6} intensities as well as the $1/e$ widths of the \ion{H}{1}
Ly$\alpha$ and \ion {O}{6} profiles. The required electron density in
the CME at $1.50~R_\odot$ was $n_e\approx 1.6\times 10^7$ ~cm$^{-3}$
and the required outflow velocity was $\approx 230$~km~s$^{-1}$, which
agrees well with with the acceleration scenario. The best fit to the
electron temperature was $T_e=9\times 10^5$~K, where we assumed oxygen
and hydrogen ionization equilibrium. Figure~3 shows the model
prediction for the \ion{H}{1} Ly$\alpha$ and \ion {O}{6}
$\lambda\lambda1032,1037$ line intensities for different $n_e$ and
$T_e$ ranging from $0.5\times10^6$~K and $1.5\times10^6$~K at
$1.98~R_\odot$. The best fit to the observed values required
$n_e\approx6.1\times 10^6$, $T_e=0.5-1\times10^6$~K, and the CME
outflow velocity ranging from 240 to 140~km~s$^{-1}$ where the higher
velocities correspond to lower electron temperature. Higher electron
temperatures (see solution for $1.5\times10^6$~K) are still marginally
within the range allowed by observations but they yield low outflow
velocities ($\sim85$~km~s$^{-1}$). To reproduce the observed line
widths at both heights we assumed $T_\perp$ to be equal to the kinetic
temperature derived from the \ion{H}{1} Ly$\alpha$ and \ion{O}{6}
$1/e$ line widths and we set $T_\perp=T_\parallel$. Coulomb collisions
are expected to be strong enough to maintain isotropic velocity
distribution in these dense enhancements.

\begin{figure}
\epsscale{1.0}
\plotone{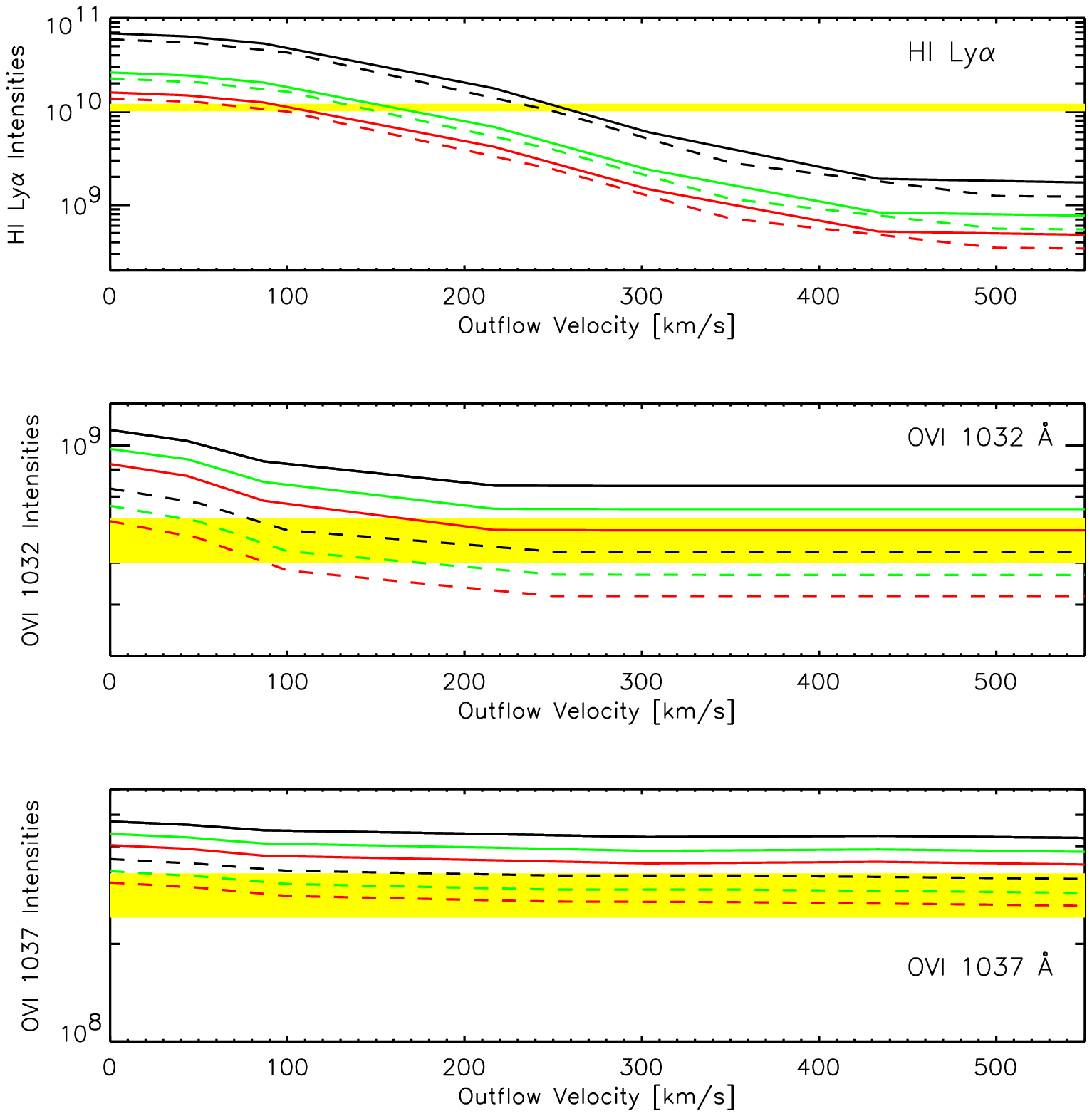}
\caption{Model \ion{H}{1} Ly$\alpha$ and \ion{O}{6}
($\lambda\lambda1032,1037$) intensities as a function of outflow
velocity for the 1999 April 15 event at $1.98~R_\odot$. The line
intensities are in photons~s$^{-1}$~cm$^{-2}$~sr$^{-1}$ and velocities
in km~s$^{-1}$. As in Figure~2, the different line styles correspond
to different electron densities: $7.0\times10^6$, solid, and
$5.9\times10^6$~cm$^{-3}$, dashed. Predictions for different electron
temperatures are marked with different colors: $T_e=0.5\times10^6$~K -
black, $T_e=1\times10^6$~K - green, and $T_e=1.5\times10^5$~K - red.
The yellow block corresponds to observationally allowed
values.}
\end{figure}

\subsection{1999 May 09 -- unstructured CME}

On 1999 May 09 UVCS recorded brightening of the emission lines at
$1.38, 1.58$ and $1.86~R_\odot$ by $15 - 28$\% for the \ion{H}{1}
Ly$\alpha$ and $25 - 50$\% for the \ion {O}{6} respectively (Table~2).
The $1/e$ widths of the \ion{H}{1} Ly$\alpha$ remained unchanged while
the \ion{O}{6} profiles appeared only slightly narrower (by
$5-10$~km~s$^{-1}$) than the pre--CME corona. At $1.58~R_\odot$,
within the uncertainties of the observed values, we obtained the best
fit for $n_e=1.2\times10^7$~cm$^{-3}$, $T_e\approx 1-1.2\times10^6$~K
and speeds of $\approx 150-115$~km~s$^{-1}$, where the higher speeds
correspond to lower electron temperatures. The model was able to
reproduce the observed values at $1.86~R_\odot$ with
$n_e=4.1\times10^6$~cm$^{-3}$, electron temperature in a range of
$T_e\approx 0.5-1\times10^6$~K and corresponding outflow speeds of
$\approx 250-150$~km~s$^{-1}$.  The ion temperature was set to
$T_\perp = T_\parallel$

\subsection{Jets or Flux Ropes}

The prevalent physical models for CMEs and jets are quite different.
MHD models of CMEs show plasma lifted by the upward motion of
transverse magnetic field lines, (e.g. Antiochos et al. 1999), with
the field lines in the form of a flux rope in many models (e.g. Gibson
\& Low 1998; Amari et al. 2000; Lin \& Forbes 2000; Manchester et
al. 2002).  Jets, on the other hand, are imagined to involve strong
heating at the injection site (e.g. Shimojo et al. 1996), but
relatively cool material may be ejected. Kahler, Reames \& Sheeley
(2001) have suggested that flare-generated energetic particles
observed at 1~AU are associated with narrow CMEs because these involve
reconnection onto open field lines.  This would suggest that narrow
CMEs are more powerful versions of jets.

In the standard CME picture, reconnection will produce some local
heating, but the models do not require heating for the bulk of the
ejected gas.  On the other hand, strong heating is observed at the
ejection site (e.g. Filippov \& Koutchmy 2002) and continued heating
is inferred from the temperatures and ionization states observed at
several solar radii (Akmal et al. 2001; Ciaravella et al. 2001).
Observations of polar coronal jets also indicate continued heating as
the plasma moves upward through the corona (Dobrzycka et al. 2000).

We have good diagnostic information for 2 CMEs, one structured and one
unstructured.  They have very similar densities and temperatures.  The
densities and temperatures are somewhat higher than those obtained by
Dobrzycka et al. (2002) for polar coronal jets, but only by a factor
of 2 or less.  CMEs show a large range of temperatures as indicated by
ions ranging from C III to Fe XXI (Raymond 2002), and densities are
comparable to those inferred for the 1999 April 15 and 1999 June 3
events (e.g. Ciaravella et al. 2001) or higher. The speeds derived for
the narrow ejections observed here are comparable to those of polar
hole jets (Dobrzycka et al. 2000, 2002), and in the low to moderate
speed range of CMEs.  There is no obvious line width or velocity
signature, such as the rotation observed in some CMEs (Ciaravella et
al. 2000; Pike \& Mason 2002). However, the acceleration of the 1999
April 15 event is similar to that observed in slow CMEs, but not
typical of jets (Wang et al. 1998, Wood et al. 1999). This strongly
indicates that the structured event is similar to larger CMEs rather
than jets.

Overall, we conclude that the UVCS observations do not definitively
place the narrow CMEs into the jet picture of reconnection onto open
field lines or the CME picture of expanding closed field structures.
If anything, their parameters are intermediate between those of jets
and of CMEs. The higher sensitivity, spatial and temporal resolution
of the proposed {\it Advanced Spectroscopic and Coronagraphic Explorer
(ASCE)} mission should either break the structured narrow CMEs down
into sinuous filaments like those observed in larger CMEs or else
reveal line intensity diagnostics indicating impulsive heating at the
injection site.  In the mean time, a brighter event observed with a
fortunate choice of observing parameters might allow the {\it SOHO}
instruments to differentiate between the jet and CME pictures.

\section{Summary}

We presented UVCS observations of five narrow CMEs. They were among 15
narrow CMEs originally selected by Gilbert et al. (2001) based on
their white light morphology and disk association. Two events (1999
March 27, April 15) were structured, i.e. in white light data they
exhibited well defined interior features, and three (1999 May 9, May
21, June 3) were unstructured, i.e. appeared featureless. In UVCS data
the events appeared as enhancements of the strongest coronal lines
\ion{H}{1} Ly$\alpha$ and \ion{O}{6} $\lambda\lambda1032,1037$. They
were $4^\circ-13^\circ$ wide and their signature was recorded at
several heights. Observations by the EIT and SXT instruments suggest
that the narrow CMEs are associated with disk activity that lasted up
to several hours. In the LASCO C2 field of view each CME structure was
visible for an extended period of time of about $1-1.5$ hour.

We derived the electron densities for several of the events from the
LASCO C2 white light observations (Table 3). They are
$1.5-3\times10^6$~cm$^{-3}$ for the 1999 April 15 and 1999 June 3
events, which is comparable to densities inferred for regular CMEs
(e.g. Ciaravella 2001). The narrow ejection on 1999 May 9 appeared to
be less dense, $4.5\times10^5$~cm$^{-3}$. 

We used a spectral line synthesis code based on the code developed by
Cranmer et al. (1999) to model plasma properties of the best observed
examples of the structured and unstructured narrow CMEs. The CMEs were
described as flux tubes located in the plane of the sky with no
contribution from the ambient corona. We modeled the best observed
examples of structured (1999 April 15) and unstructured (1999 May 9)
narrow CMEs. We fit the \ion{H}{1} Ly$\alpha$ and \ion{O}{6}
$\lambda\lambda1032,1037$ intensities, the $1/e$ widths as well as the
\ion{O}{6} line ratios at each selected height.

For the April 15 event we obtained the best fit for $n_e\approx
1.6\times10^7$~cm$^{-3}$, $T_e\approx9\times10^5$~K and outflow
velocity of $\approx230$~km~s$^{-1}$ at $1.50~R_\odot$, which agrees
with the acceleration scenario, and $n_e\approx
6.1\times10^6$~cm$^{-3}$, $T_e\approx0.5-1\times10^6$~K with the CME
speed from 240 to 140~km~s$^{-1}$ (where the higher speed
corresponding to lower $T_e$) at $1.98~R_\odot$.

The best fit to the May 9 observations at $1.58~R_\odot$ required
$n_e=1.2\times10^7$~cm$^{-3}$, $T_e\approx 1-1.2\times10^6$~K and
speeds of $\approx 150-115$~km~s$^{-1}$, where the higher speeds
correspond to lower electron temperatures. At $1.86~R_\odot$ the model
was able to reproduce the observed values with
$n_e=4.1\times10^6$~cm$^{-3}$, electron temperature in the range of
$T_e\approx 0.5-1\times10^6$~K and corresponding outflow speeds of
$\approx 250-150$~km~s$^{-1}$, assuming an ion temperature equal to
$T_\perp = T_\parallel$. Gilbert et al. (2001) found the speed of the
CME to be 318~km~s$^{-1}$ within the C2 field of view.

The derived plasma parameters for the structured and unstructured
narrow CMEs look very similar. Compared to the polar coronal jets they
have comparable speeds, higher densities and temperatures but only by
a factor of 2 or less. With respect to the CMEs showing a large range
of densities, temperatures and outflow speeds the narrow ejections'
plasma parameters look comparable. We did not see any obvious line
width and velocity signature, such as rotation characteristic for some
CMEs. However, the acceleration of the 1999 April 15 event is similar
to that observed in slow CMEs, but not typical of jets (Wang et
al. 1998, Wood et al. 1999).

Overall, we found that the UVCS observations do not definitively place
the narrow CMEs into the jet picture of reconnection onto open field
lines or the CME picture of expanding closed field structures.
Additional observations of brighter events with more suitable
observing parameters or with more sensitive instruments might allow us
to differentiate between the jet and CME scenarios.

\acknowledgments

The authors thank Steven R. Cranmer for discussion and comments about
the spectral line synthesis model. We acknowledge use of the CME
catalog generated and maintained by the Center for Solar Physics and
Space Weather, The Catholic University of America in cooperation with
the Naval Research Laboratory and NASA. SOHO is a project of
international cooperation between ESA and NASA. This work is supported
by the National Aeronautics and Space Administration under grant
NAG5--11420 to the Smithsonian Astrophysical Observatory, by Agenzia
Spaziale Italiana, and by the ESA PRODEX program (Swiss contribution).

\clearpage
\begin{deluxetable}{lcccccccc}
\tablewidth{0pt}
\tabletypesize{\scriptsize}
\tablecaption{UVCS/{\it SOHO} Observations of Narrow CMEs}
\tablenum{1}
\tablehead{ 
\multicolumn{1}{c}{} & 
\multicolumn{4}{c}{UVCS} &
\multicolumn{4}{c}{LASCO C2} \\
\cline{2-5}
\cline{6-9}
\colhead{Date} &
\multicolumn{1}{c}{$\rho$\tablenotemark{a}} & \multicolumn{1}{c}{P.A.} &
\multicolumn{1}{c}{$\Delta\phi$\tablenotemark{b}} &
\multicolumn{1}{c}{$t_{app}\tablenotemark{c}$} &
\multicolumn{1}{c}{P.A.} &
\multicolumn{1}{c}{$\Delta\phi$\tablenotemark{b}} &
\multicolumn{1}{c}{$t_{app}$\tablenotemark{c}}  &
\multicolumn{1}{c}{Speed} \\
\multicolumn{1}{c}{} &
\multicolumn{1}{c}{$R_\odot$} & \multicolumn{1}{c}{deg} &
\multicolumn{1}{c}{deg} & \multicolumn{1}{c}{UT} &
\multicolumn{1}{c}{deg} & \multicolumn{1}{c}{deg} &
\multicolumn{1}{c}{UT} &
\multicolumn{1}{c}{km~s$^{-1}$} }
\startdata 
03/27/99&1.49&314& 13   &16:02  &310&12 &16:54&475  \\
        &1.85&312& 10   &16:29 &   &   &     &      \\
04/15/99&1.38&109& 7    & 07:33 &102& 10&07:54&523   \\ 
        &1.50&108&7     & 07:40 &   &   &     &        \\
        &1.62&107&7     & 07:47 &   &   &     &        \\
        &1.74&107&7     & 07:59 &   &   &     &        \\
        &1.98&106&6     & 08:10 &   &   &     &        \\
        &2.33&105&6     & 08:24 &   &   &     &        \\
        &2.88&104&6     & 08:42 &   &   &     &        \\ 
        &3.62&102&5     & 09:27 &   &   &     &        \\
05/09/99&1.38&100&4     & 19:17 &100& 4 &19:30&318    \\
        &1.58&100&4     & 19:24 &   &   &     &        \\
        &1.86&100&4     & 19:32 &   &   &     &        \\
        &2.15&100&4     & 19:43 &   &   &     &        \\
05/21/99&3.62&289&      & 02:08 &290& 8 &02:06&241     \\
06/03/99&2.88& 86& 4    & 10:15 & 90& 4 &07:26&308    \\
        &     &   &      &       &   &   &     &      \\
\enddata

\tablenotetext{a}{$\rho$ is the projected height measured from Sun center}
\tablenotetext{b}{$\Delta\phi$ is the angular width of the structure}
\tablenotetext{c}{$t_{app}$ is the time of appearance of the narrow
CME in UVCS slit and LASCO C2 field of view as in Gilbert et al. (
2001)}
\end{deluxetable}


\begin{deluxetable}{lccccccccccccc}
\rotate
\tablewidth{0pt}
\tabletypesize{\scriptsize}
\tablecaption{Emission Line Properties }
\tablenum{2}
\tablehead{ 
\multicolumn{2}{c}{} & 
\multicolumn{4}{c}{H~I Ly$\alpha$} &
\multicolumn{4}{c}{O~VI $\lambda1032$} &
\multicolumn{4}{c}{O~VI $\lambda1037$} \\
\cline{3-6}
\cline{7-10}
\cline{11-14}\\[-1.6ex]
\colhead{Date} &
\multicolumn{1}{c}{$\rho$} & 
\multicolumn{1}{c}{$I$\tablenotemark{a}} &
\multicolumn{1}{c}{$I/I_{b}$} &
\multicolumn{1}{c}{$V_{1/e}$\tablenotemark{b}} & 
\multicolumn{1}{c}{$V_{1/e}/V_{1/e,b}$} & 
\multicolumn{1}{c}{$I$\tablenotemark{a}} &
\multicolumn{1}{c}{$I/I_{b}$} &
\multicolumn{1}{c}{$V_{1/e}$\tablenotemark{b}} & 
\multicolumn{1}{c}{$V_{1/e}/V_{1/e,b}$} & 
\multicolumn{1}{c}{$I$\tablenotemark{a}} &
\multicolumn{1}{c}{$I/I_{b}$} &
\multicolumn{1}{c}{$V_{1/e}$\tablenotemark{b}} & 
\multicolumn{1}{c}{$V_{1/e}/V_{1/e,b}$} }
\startdata 
03/27/99&1.49&1.0e11&0.32&145&1 &3.0e9&0.16 & 65& 1   &1.3e9&0.2& 70&0.93\\
        &1.85&2.7e10&0.38&150&1 &1.3e9&0.60 & 60& 1.1 & 7.1e8&1.0&  &    \\
04/15/99&1.38&4.5e10&0.25&101&0.65&7.8e9 &1.15& 74&1.06&3.4e9&1.4 &74&1.05\\
        &1.50&3.6e10&0.35&149&0.96&4.0e9 &1.02& 69&0.98&2.1e9&1.5 &70&0.97\\
        &1.62&2.2e10&0.37&138&0.92&2.8e9 &1.75& 72&0.90&1.1e9&1.92&73&0.85\\
        &1.74&1.7e10&0.41&160&1.02&1.5e9 &2.29&77&0.97&7.0e8&$\sim3$&72&1.0\\
        &1.98&1.1e10 &0.50&135&0.87&7.0e8 &3.5 &82 &1.00&3.1e8&    &82&1.0 \\
        &2.33&4.1e9 &0.41&134&0.81&2.2e8 &$\sim4$& &   &$\sim1e8$& &  &    \\
        &2.88&8.7e8 &0.27&165&1.0 &3.3e7 &2.4 &   &    &1.6e7&    &   &   \\
        &3.62&3.9e8 &0.35&   &    &1.3e7 &    &   &    &     &    &   &   \\
05/09/99&1.38&1.1e11&0.26&142&1.0 &9.2e9 &0.35& 62&0.87&4.5e9&0.5 &70&1.0 \\
        &1.58&4.8e10&0.28&145&1.0 &2.5e9 &0.30& 67&0.84&1.2e9&0.4 &75&1.0 \\
        &1.86&1.0e10 &0.17&145&$\sim1$&3.1e8&0.20&70&0.80&1.5e8&0.25&  &   \\
        &2.15&      &    &   &    &$\sim3$e8&0.6 &   & &$\sim1$e8&0.6& &  \\
05/21/99&3.62&      &    &   &    &      &0.5 &   &    &     &    &   &   \\
06/03/99&2.88&8.6e8 &0.12&   &    &6.9e7 &0.75&   &    &2.7e7&0.96&   &   \\
        &    &      &    &   &    &      &    &   &    &     &    &   &   \\
\enddata
\tablenotetext{a}{Intensities correspond to the observed integrated
profiles with subtracted pre- or post-CME intensities ($I_b$). The units are
photons~s$^{-1}$~cm$^{-2}$~sr$^ {-1}$}

\tablenotetext{b}{The $1/e$ widths are in km~s$^{-1}$}
\end{deluxetable}

\begin{deluxetable}{lcccc}
\tablewidth{0pt}
\tabletypesize{\scriptsize}
\tablecaption{Electron Densities of Narrow CMEs}
\tablenum{3}
\tablehead{ 
\colhead{Date} &
\colhead{Time} &
\colhead{Aver. height} &
\colhead{$\Delta\phi$\tablenotemark{a}} &
\colhead{$N_e$} \\
\colhead{} &
\colhead{UT} &
\colhead{$R_\odot$} &
\colhead{deg} &
\colhead{$cm^{-3}$} }
\startdata 
04/15/99 & 07:54 & 2.77  & 1.9 &  3.0e6  \\
         & 08:06 & 2.88  & 2.2 &  3.0e6  \\
         & 08:30 & 3.30  & 2.2 &  2.7e6  \\
         & 08:54 & 3.53  & 4.0 &  1.3e6  \\
         & 09:06 & 3.65  & 5.0 &  1.1e6  \\
         & 09:30 & 3.92  & 5.5 &  1.1e6  \\
05/09/99 & 19:28 & 2.91  & 5.2 &  4.4e5  \\
         & 19:51 & 3.37  & 4.5 &  4.5e5  \\
         & 20:27 & 3.87  & 4.4 &  4.3e5  \\
06/03/99 & 07:26 & 2.45  & 2.8 &  1.4e6  \\
         & 07:50 & 2.85  & 4.7 &  1.3e6  \\
         & 08:06 & 3.14  & 5.3 &  8.8e5  \\
         & 08:26 & 3.42  & 4.8 &  8.0e5  \\
         &       &       &     &         \\
\enddata
\tablenotetext{a}{$\Delta\phi$ is the angular width of the structure}
\end{deluxetable}

\clearpage

\end{document}